\documentclass[12pt]{article}
 \usepackage{epsfig}
 \def\be{\begin{equation}}
 \def\ee{\end{equation}}
 \def\bea{\begin{eqnarray}}
 \def\eea{\end{eqnarray}}
 \usepackage{graphicx}% Include figure files

 \catcode`\@=11
 \def\lsim{\mathrel{\mathpalette\@versim<}}
 \def\gsim{\mathrel{\mathpalette\@versim>}}
 \def\@versim#1#2{\vcenter{\offinterlineskip
 \ialign{$\m@th#1\hfil##\hfil$\crcr#2\crcr\sim\crcr } }}
 \catcode`\@=12

 \catcode`@=12
\begin{document}
 \thispagestyle{empty}
 \begin{flushright}
 UCRHEP-T619\\
 Mar 2022\
 \end{flushright}
 \vspace{0.6in}
 \begin{center}
 {\LARGE \bf Scotogenic $A_5 \to A_4$ Dirac Neutrinos\\
with Freeze-In Dark Matter\\}
 \vspace{1.5in}
 {\bf Ernest Ma\\}
 \vspace{0.1in}
{\sl Department of Physics and Astronomy,\\ 
University of California, Riverside, California 92521, USA\\}
\end{center}
 \vspace{1.2in}

\begin{abstract}\
Radiative Dirac neutrino masses and their mixing are linked to   
dark matter through the non-Abelian discrete symmetry $A_5$ of the 
4-dimensional pentatope, softly broken to $A_4$ of the 3-dimensional 
tetrahedron.  This unifying understanding of neutrino family structure from 
dark matter is made possible through the interplay of gauge symmetry, 
renormalizable Lagrangian field theory, and softly broken discrete symmetries.
Dark neutral fermions are produced through Higgs decay.
\end{abstract}

\newpage
\baselineskip 24pt
\noindent \underline{\it Introduction}~:~ 
Two fundamental issues in particle physics and astroparticle physics are 
neutrinos~\cite{g16} and dark matter~\cite{y17}.  An important development 
in recent years is the notion that they are intrinsically related, i.e. 
the origin of neutrino masses is the existence of dark matter.  This is 
simply accomplished in the scotogenic model~\cite{m06}, where radiative 
Majorana neutrino masses are generated in one loop with the internal 
particles belonging to the dark sector.  Numerous variations and studies 
of this basic premise have appeared in the past 16 years.  More recently, 
the notion that neutrinos are Dirac fermions~\cite{mp17} has received more 
attention, and the scotogenic mechanism may also be applied~\cite{m20}.

Since neutrinos come in three families, their $3 \times 3$ mixing matrix 
with charged leptons could be a hint to a symmetry yet to be discovered 
among them.  In the context of a renormalizable Lagrangian field theory 
subject to the $SU(3) \times SU(2) \times U(1)$ gauge symmetry of the 
Standard Model (SM), this poses a conundrum.  How is it that a symmetry 
may govern mixing and yet all lepton masses are so different?  This 
problem was solved 21 years ago~\cite{mr01} using the non-Abelian discrete 
symmetry $A_4$ and its breaking, which allow three arbitrary charged-lepton 
masses, and yet maintain a specific pattern for the neutrino mass matrix, 
which results in tribimaximal~\cite{hps02} or cobimaximal 
mixing~\cite{bmv03,gl04}, close to what is observed experimentally.

In the original scotogenic model for Majorana neutrinos, the dark sector 
is distinguished by a dark parity, which may be derived from lepton 
parity~\cite{m15}.  For Dirac neutrinos, lepton number $L$ is considered 
instead.  To prevent neutrinos from obtaining tree-level Dirac masses, 
a lepton family symmetry is used, specifically the non-Abelian discrete 
symmetry $A_5$~\cite{es09}.  With the chosen fermion and scalar multiplets 
belonging to its irreducible representations, and the soft breaking of 
$A_5 \to A_4$, scotogenic Dirac neutrino masses are obtained, with dark 
number $D = L-2j$~\cite{m20}, where $j$ is the intrinsic spin of the 
particle.  The conservation of $L$ thus implies the conservation of $D$.

\noindent \underline{\it Tetrahedron and Pentatope}~:~ 
The tetrahedron is the simplest perfect geometric solid in three dimensions. 
In Cartesian coordinates, its four vertices may be simply put at the 
positions
\begin{equation}
(1,0,0),~~~(0,1,0),~~~(0,0,1),~~~(1,1,1).
\end{equation}
Each combination of three points forms an equilateral triangle with side 
$\sqrt{2}$, and there are four such triangles.  The pentatope is the 
simplest perfect geometric solid in four dimensions. In Cartesian 
coordinates, its five vertices may be simply put at the positions
\begin{equation}
(2,0,0,0),~~~(0,2,0,0),~~~(0,0,2,0),~~~(0,0,0,2),~~~
(\varphi,\varphi,\varphi,\varphi),
\end{equation}
where $\varphi = (1+\sqrt{5})/2 = 1.618$ is the golden ratio. Each combination 
of four points forms a tetrahedron with side $2\sqrt{2}$, and there are 
five such tetrahedrons.

The tetrahedron is invariant under $A_4$, the group of the even permutation 
of four objects.  It has 12 elements and 4 irreducible representations, i.e. 
\begin{equation}
\underline{1},~~~\underline{1}',~~~\underline{1}'',~~~\underline{3}.
\end{equation}
The pentatope is invariant under $A_5$, the group of the even permutation 
of five objects.  It has 60 elements and 5 irreducible representations, i.e. 
\begin{equation}
\underline{1},~~~\underline{3},~~~\underline{3}',~~~\underline{4},
~~~\underline{5}.
\end{equation}
The multiplication rules of these $A_5$ representations are~\cite{es09}
\begin{eqnarray}
3 \times 3  &=& 1 + 3 + 5, \\ 
3' \times 3' &=& 1 + 3' + 5, \\ 
3 \times 3' &=& 4 + 5, \\ 
3 \times 4 &=& 3' + 4 + 5, \\ 
3' \times 4 &=& 3 + 4 + 5, \\ 
3 \times 5 &=& 3 + 3' + 4 + 5, \\ 
3' \times 5 &=& 3 + 3' + 4 + 5, \\ 
4 \times 4 &=& 1 + 3 + 3' + 4 + 5, \\ 
4 \times 5 &=& 3 + 3' + 4 + 5 + 5, \\ 
5 \times 5 &=& 1 + 3 + 3' + 4 + 4 + 5 + 5. 
\end{eqnarray}
Since $A_4$ is a subgroup of $A_5$, the decompositions of the latter 
represenations to the former are~\cite{es09}
\begin{equation}
1 \sim 1,~~~3 \sim 3,~~~3' \sim 3,~~~4 \sim 3 + 1,~~~5 \sim 3 + 1' + 1''.
\end{equation}
With the choice of particle content of the model to be described under $A_5$, 
the above properties of $A_5 \to A_4$ in the context of a renormalizable 
Lagrangian gauge theory will result in the radiative generation of neutrino 
masses  with a realistic family structure. 

\noindent \underline{\it Model}~:~ 
The idea of this model and its implementation are both very simple.  
\begin{table}[tbh]
\centering
\begin{tabular}{|c|c|c|c|c|c|}
\hline
fermion/scalar & $SU(2)_L \times U(1)_Y$ & $A_5$ & $A_4$ & $L$ & $D=L-2j$ \\
\hline
$L_L=(\nu,l)_L$ & $(2,-1/2)$ & $3$ & $3$ & $1$ & 0 \\ 
$l_R$ & $(1,-1)$ & $3$ & $3$ & $1$ & 0 \\ 
$\nu_R$ & $(1,0)$ & $3'$ & 3 & 1 & $0$ \\ 
\hline
$E_{L,R}$ & $(1,-1)$ & $3$ & $3$ & $1$ & 0 \\ 
\hline
$\Phi=(\phi^+,\phi^0)$ & $(2,1/2)$ & $1$ & $1$ & $0$ & 0 \\ 
\hline
$N_{L,R}$ & $(1,0)$ & $4$ & $3,1$ & $0$ & $-1$  \\ 
\hline
$\eta=(\eta^0,\eta^-)$ & $(2,-1/2)$ & $3'$ & $3$ & $1$ & 1 \\ 
$\zeta^0$ & $(1,0)$ & $3$ & $3$ & $1$ & 1 \\ 
\hline
$\zeta^-$ & $(1,-1)$ & $3'$ & $3$ & $1$ & 1 \\ 
\hline
\end{tabular}
\caption{Fermions and scalars in the $A_5 \to A_4$ model.}
\end{table}
The three copies of SM charged-lepton doublets $L_L=(\nu,l)_L$ and singlets 
$l_R$ transform each as $\underline{3}$ under $A_5$, whereas the three 
neutral singlets $\nu_R$ transform as $\underline{3}'$.  There is one Higgs 
doublet $(\phi^+,\phi^0)$ as in the SM, transforming as $\underline{1}$ 
under $A_5$.  The charged-lepton masses are obtained from heavy fermion 
singlets $E_{L,R} \sim \underline{3}$ in a seesaw manner.  For radiative 
Dirac neutrino masses, four neutral Dirac fermion singlets 
$N \sim \underline{4}$ are added, with three scalar doublets 
$(\eta^0,\eta^-) \sim \underline{3}'$ and three scalar singlets 
$\zeta^0 \sim \underline{3}$, as shown in Table 1.  There are also three 
charged scalar singlets $\zeta^- \sim \underline{3}'$ to allow the decay of  
the Higgs boson to $\bar{N}N$.  

Whereas $\bar{l}_R \Phi^\dagger L_L$ and $\bar{E}_R \Phi^\dagger L_L$ terms 
($3 \times 1 \times 3$) are allowed, 
$\bar{\nu}_R \tilde{\Phi}^\dagger L_L$ ($3' \times 1 \times 3$) is forbidden 
by $A_5$ and $\bar{N}_R \tilde{\Phi}^\dagger L_L$ ($4 \times 1 \times 3$) 
is forbidden by both $A_5$ and $L$.  Hence the neutral singlet fermions 
$\nu_R$ are not tree-level Dirac mass partners to the SM doublet neutrinos 
$\nu_L$ even though they both have $L=1$.  On the other hand, 
$\bar{L}_L \eta N_R$ ($3 \times 3' \times 4$) and $\bar{\nu}_R \zeta^0 N_L$ 
($3' \times 3 \times 4$) are allowed.  As for the $\Phi \eta \bar{\zeta}^0$ 
term, it transforms as $1 \times 3' \times 3$ which is not invariant under 
$A_5$, but it is a dimension-three soft term, so it may be chosen to break 
$A_5 \to A_4$ in which case $1 \times 3 \times 3$ is invariant under $A_4$. 
As a result, $\nu_L$ is linked to $\nu_R$ in one loop.

\noindent \underline{\it Charged Leptons}~:~ 
Under $A_5$, the $6 \times 6$ mass matrix linking $(l,E)_L$ to $(l,E)_R$ is 
given by
\begin{equation}
{\cal M} = \pmatrix{ 0 & M_{l E} \cr M_{E l} & M_{E E}},
\end{equation}
where the $3 \times 3$ entries $M_{l E}$, $M_{E l}$, and $M_{E E}$ 
are all proportional to the identity, with $M_{l E}$ coming from the 
vacuum expectation value $v$ of the SM Higgs doublet $\Phi$.  Since 
the $E_L l_R$ coupling is a dimension-three soft term, it is assumed to 
break $A_5$ in a way compatible with $A_4$ using the procedure of 
Ref.~\cite{m06-1}, i.e.
\begin{equation}
M_{E l} = \pmatrix{h_1 v' & h_2 v'' & h_3 v'' \cr h_3 v'' & h_1 v' 
& h_2 v'' \cr h_2 v'' & h_3 v'' & h_1 v'},
\end{equation}
which is obtained from $E \sim 3$ and $l_R \sim 3$ and a gauge singlet flavon 
$\sim 3$ under $A_4$.  The magic of this matrix is that it decomposes 
to~\cite{m06-1} 
\begin{equation}
U_L \pmatrix{h_1 v' + (h_2+h_3)v'' & 0 & 0 \cr 0 & h_1 v' + (h_2 \omega + 
h_3 \omega^2) v'' & 0 \cr 0 & 0 & h_1 v' + (h_2 \omega^2 + h_3 \omega) v''} 
U_R^\dagger,
\end{equation}
where
\begin{equation}
U_L = U_R = U_\omega = {1 \over \sqrt{3}} \pmatrix{ 1 & 1 & 1 \cr 1 & 
\omega & \omega^2 \cr 1 & \omega^2 & \omega},
\end{equation}
with $\omega = \exp(2 \pi i/3) = -1/2 + i \sqrt{3}/2$, is the well-known 
transformation matrix studied in numerous $A_4$ models.  Assuming that the 
singlet $3 \times 3$ $M_{E E}$ masses to be much heavier than $M_{l E}$ 
and $M_{E l}$, the $3 \times 3$ charged-lepton mass matrix is given by
\begin{equation}
{M}_{l l} = M_{l E} M^{-1}_{E E} M_{E l},
\end{equation}
which preserves the form of Eq.~(18).  Hence the charged-lepton mass 
matrix has three independent eigenvalues, which may be chosen to be 
$m_e,m_\mu,m_\tau$ and yet a definite mixing matrix $U_\omega$ of Eq.~(19) 
is obtained relative to the neutrino mass matrix, as in the original 
$A_4$ model~\cite{mr01}.

\noindent \underline{\it Scotogenic Neutrino Masses}~:~ 
With the $A_5$ assignments of Table~1, Dirac neutrino masses are 
 \begin{figure}[htb]
 \vspace*{-5cm}
 \hspace*{-3cm}
 \includegraphics[scale=1.0]{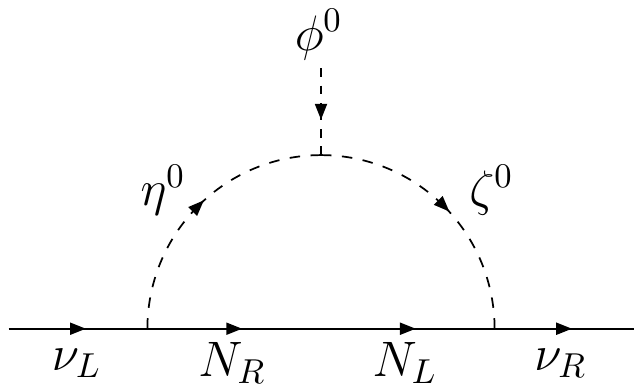}
 \vspace*{-21.5cm}
 \caption{One-loop scotogenic Dirac neutrino mass.}
 \end{figure}
generated in one loop as shown in Fig.~1.  To compute this diagram, the 
first step is to note that the $A_5 \to A_4$ breaking trilinear coupling 
$\eta^0 \phi^0 \bar{\zeta}^0$ with $\langle \phi^0 \rangle = v$ means  
that $\eta^0$ mixes with $\zeta^0$ in a $2 \times 2$ mass-squared matrix. 
Hence the two mass eigenstates $\psi^0_{1,2}$ with $m_{1,2}$ are 
\begin{equation}
\psi^0_1 = \eta^0 \cos \theta - \zeta^0 \sin \theta, ~~~  
\psi^0_2 = \eta^0 \sin \theta + \zeta^0 \cos \theta.
\end{equation}  
The second step is to note that the $\bar{\nu}_L \eta^0 N_R$ and 
$\bar{\nu}_R \zeta^0 N_L$ couplings come from the decomposition 
of $3 \times 3' \to 4$.  Let $(a_1,a_2,a_3) \sim 3$, $(b_1,b_2,b_3) \sim 3'$, 
then the four components of $N$ are given by~\cite{es09}
\begin{equation}
N = {1 \over \sqrt{3}} \pmatrix{\varphi^{-1} a_3 b_2 - \varphi a_1 b_3 \cr 
\varphi a_3 b_1 + \varphi^{-1} a_2 b_3 \cr -\varphi^{-1} a_1 b_1 + 
\varphi a_2 b_2 \cr a_2 b_1 - a_1 b_2 + a_3 b_3}.
\end{equation}
In the $A_5$ limit, all components of $N$ have the same mass, but since 
the $4 \times 4$ ${\cal M}_N$ mass matrix is a soft term, it is allowed 
to break $A_5$.  The form of ${\cal M}_\nu$ may then be chosen as in 
previous $A_4$ models to obtain cobimaximal mixing for example.  This in 
turn would imply a family structure for the dark $N$ fourplet.

The radiative Dirac neutrino mass matrix is
\begin{equation}
({\cal M}_\nu)_{ij} = {\sin \theta \cos \theta \over 16 \pi^2} \sum_{k,k',a} 
f^L_{ika} f^\phi_{k'k4} f^R_{k'ja} M_a [F(m^2_2,M^2_a) - F(m^2_1,M^2_a)],
\end{equation}
where $M_a$ ($a=1,2,3,4$) are the masses of the four $N$ fermions.  
The function $F$ is given by
\begin{equation}
F(x,y) = {x \ln (x/y) \over x-y}.
\end{equation}
The $f_{ika}$ couplings are as defined by Eq.~(22), with $i=1,2,3$ from 
$\underline{3}$, $k=1,2,3$ from $\underline{3}'$, and $a=1,2,3,4$ from 
$\underline{4}$.  If all $M_a$ are equal, all three Dirac neutrinos would 
have the same mass, as expected.

It is now assumed that the $4 \times 4$ $M_N$ mass matrix is not invariant 
under $A_5$.  Note first that $\eta_{1,2,3}, \zeta_{1,2,3} \sim 3$,  
$N_{1,2,3} \sim 3$ and $N_4 \sim 1$ under $A_4$.  Assume then that $N_4$ 
is very much heavier than all other masses.  As for $N_{1,2,3}$, they form 
a mass matrix $M_{ab}$ which softly breaks $A_4$ with all entries much 
lighter than $m_{1,2}$.  This reduces ${\cal M}_\nu$ 
to the form first recognized in Ref.~\cite{m12}, i.e.
\begin{equation}
({\cal M}_\nu)_{ij} = {\sin \theta \cos \theta \ln (m_2^2/m_1^2) \over 16 
\pi^2} \sum_{k,k',a,b} f^L_{ika} f^\phi_{k'k4} f^R_{k'jb} M_{ab},
\end{equation}
where $M_{ab}$ links $N_R$ to $N_L$.  Using Eq.~(22), the above is then 
proportional to
\begin{equation}
\tilde{\cal M} = \pmatrix{ \varphi^2 M_{12} & M_{11}+M_{33} & \varphi^{-2} 
M_{32} \cr M_{22}+M_{33} & \varphi^{-2} M_{21} & \varphi^{2} M_{31} 
\cr \varphi^{-2} M_{13} & \varphi^2 M_{23} & M_{11}+M_{22}},
\end{equation}
where $N_{R1}$ has been redefined with a minus sign.  If 
$M_{11}=M_{22}=M_{33}$ and all $M_{ij} =0$ with $i \neq j$, this 
reduces to the $A_5$ coupling of $3 \times 3'$ to the fourth component 
of $\underline{4}$, as expected.

To achieve cobimaximal mixing~\cite{m21}, i.e. $\theta_{13} \neq 0$, 
$\sin^2 \theta_{23} = 1/2$, and $\delta_{CP} = \pi/2,3\pi/2$, the above 
mass matrix should be diagonalized by an orthogonal matrix ${\cal O}$ 
on the left, so that the $3 \times 3$ neutrino mixing matrix becomes
\begin{equation}
U_{l\nu} = U_\omega^\dagger {\cal O},
\end{equation}
where $U_\omega$ comes from Eq.~(19).  It was shown 22 years ago~\cite{fmty00} 
that this results automatically in cobimaximal mixing.

An equivalent formulation~\cite{gl04} for Majorana neutrinos is to consider 
the neutrino mass matrix in the basis of diagonal charged-lepton masses, i.e.
\begin{equation}
U_\omega^\dagger {\cal M}_\nu (U_\omega^\dagger)^T = \pmatrix{A & C & E^* 
\cr C & D^* & B \cr E^* & B & F}.
\end{equation}
The conditions for cobimaximal mixing are then~\cite{m16}
\begin{equation}
E=C, ~~~ F=D, ~~~ A,B~{\rm real}.
\end{equation}
However for Dirac neutrinos, there are no unique conditions, because the 
$3 \times 3$ matrix diagonalizing ${\cal M}_\nu$ on the right is not 
constrained.  Nevertheless, a suggestive form is~\cite{m21}
\begin{equation}
{\cal M}_D = \pmatrix{a & c & c^* \cr d & b & e \cr d^* & e^* & b^*},
\end{equation}
where  $a$ is real.  Assuming $\tilde{M}$ of Eq.~(26) to be real, then 
$U_\omega^\dagger \tilde{M} U_\omega$ is automatically of this form.

\noindent \underline{\it Family Structure of Neutrinos and Dark Matter}~:~
Until 2012, the data were consistent with tribimaximal mixing, i.e. 
$\theta_{13}=0$, $\sin^2 \theta_{23}=1/2$, and $\sin^2 \theta_{12} = 1/3$.  
Now they are closer to cobimaximal mixing with $\delta_{CP}=3\pi/2$.  
The present world averages are~\cite{pdg20}
\begin{eqnarray}
&& \sin^2 \theta_{13} = (2.20 \pm 0.07) \times 10^{-2}, ~~~ 
\sin^2 \theta_{12} = 0.307 \pm 0.013, \\  
&& \sin^2 \theta_{23} = 0.546 \pm 0.021~({\rm Normal~order}), \\ 
&& \sin^2 \theta_{23} = 0.539 \pm 0.022~({\rm Inverted~order}), \\ 
&& \delta_{CP} = 1.36~(+0.20/-0.16)~\pi, ~~~  
\Delta m^2_{21} = (7.53 \pm 0.18) \times 10^{-5}~{\rm eV}^2, \\  
&& \Delta m^2_{32} = (2.453 \pm 0.033) \times 
10^{-3}~{\rm eV}^2~({\rm Normal~order}), \\ 
&& \Delta m^2_{32} = (-2.536 \pm 0.034) \times 
10^{-3}~{\rm eV}^2~({\rm Inverted~order}).
\end{eqnarray} 

With Eq.~(26), a connection is predicted between neutrinos and dark matter. 
As an example, consider the case of cobimaximal mixing, with 
\begin{equation}
U_{l \nu} = \pmatrix{c_{12} c_{13} & s_{12} c_{13} & s_{13} \cr 
(s_{12} - i c_{12} s_{13})/\sqrt{2} & (-c_{12} - i s_{12} s_{13})/\sqrt{2} & 
i c_{13}/\sqrt{2} \cr (s_{12} + i c_{12} s_{13})/\sqrt{2} & 
(-c_{12} + i s_{12} s_{13})/\sqrt{2} & -i c_{13}/\sqrt{2}}.
\end{equation}
Using Eq.~(27), the orthogonal matrix ${\cal O}$ which diagonalizes Eq.~(26) 
on the left is then
\begin{equation}
{\cal O} = {1 \over \sqrt{3}} \pmatrix{c_{12} c_{13} + \sqrt{2} s_{12} & 
s_{12} c_{13} - \sqrt{2} c_{12} & s_{13} \cr c_{12} c_{13} - s_{12}/\sqrt{2} 
+\sqrt{3/2} c_{12} s_{13} & s_{12} c_{13} + c_{12}/\sqrt{2} + \sqrt{3/2} 
c_{12} s_{13} & s_{13} - \sqrt{3/2} c_{13} \cr c_{12} c_{13} - s_{12}/\sqrt{2} 
- \sqrt{3/2} c_{12} s_{13} & s_{12} c_{13} + c_{12}/\sqrt{2} - \sqrt{3/2} 
c_{12} s_{13} & s_{13} + \sqrt{3/2} c_{13}}  
\end{equation}
For the central values
\begin{equation}
s_{13} = 0.148, ~~~ c_{13} = 0.989, ~~~ s_{12} = 0.554, ~~~ c_{12} = 0.832,
\end{equation}
the orthogonal matrix becomes
\begin{equation}
{\cal O} = \pmatrix{0.927 & -0.363 & 0.085 \cr 0.336 & 0.714 & -0.614 \cr 
0.162 & 0.598 & 0.785},
\end{equation} 
which diagonalizes ${\tilde{M}\tilde{M}^T}$.
In terms of the neutrino mass eigenvalues $m^2_{1,2,3}$, the $M_{ij}$ 
entries of Eq.~(26) are then related to ${\cal O}$.

There are nine parameters in $\tilde{M}$.  Assumming the three conditions 
\begin{equation}
\varphi^2 \pmatrix{M_{12} \cr M_{31} \cr M_{23}} = \varphi^{-2} 
\pmatrix{M_{21} \cr M_{13} \cr M_{32}},
\end{equation}
then $\tilde{M}$ with an overall scale factor becomes of the form
\begin{equation}
\tilde{M} = \pmatrix{y_1 & x_1 & y_3 \cr x_2 & y_1 & y_2 \cr y_2 & y_3 & x_3},
\end{equation}
resulting in
\begin{equation}
\pmatrix{x_1^2+y_1^2+y_3^2 \cr x_2^2+y_1^2+y_2^2 \cr x_3^2+y_2^2+y_3^2} = 
\pmatrix{0.859 & 0.113 & 0.026 \cr 0.132 & 0.510 & 0.358 \cr 0.007 & 0.377 & 
0.616} \pmatrix{m_1^2 \cr m_2^2 \cr m_3^2},
\end{equation}
and
\begin{equation}
\pmatrix{y_1(x_1+x_2)+y_2y_3 \cr y_3(x_1+x_3)+y_1y_2 \cr y_2(x_2+x_3)+y_1y_3} 
= \pmatrix{-0.337 & 0.240 & 0.097 \cr 0.079 & -0.206 & 0.127 \cr -0.031 
& -0.438 & 0.469} \pmatrix{m_1^2 \cr m_2^2 \cr m_3^2}.
\end{equation}
Assuming normal ordering of neutrino masses, the above six equations may 
be solved simply for $y_1=0$, resulting in
\begin{eqnarray}
&& x_1 = 0.597 \times 10^{-2}~{\rm eV}, ~~~ 
x_2 = 1.334 \times 10^{-2}~{\rm eV}, ~~~ 
x_3 = 2.711 \times 10^{-2}~{\rm eV}, \\  
&& y_2 = 2.850 \times 10^{-2}~{\rm eV}, ~~~ 
y_3 = 0.924 \times 10^{-2}~{\rm eV}, ~~~ 
m_1 = 0.684 \times 10^{-2}~{\rm eV}.
\end{eqnarray}
This implies $m_2 = 0.011$ eV and $m_3 = 0.051$ eV, for a sum of three 
neutrino masses = 0.07 eV, comfortably below the astrophysical bound 
of 0.15 eV. 

\noindent \underline{\it Production of Dark Matter}~:~
In Fig.~1, lepton number $L$ is conserved with $L=1$ for $\nu,\eta^0,\zeta^0$ 
and $L=0$ for $N$.  Defining $D=L-2j$ as shown in Table~1, dark number is 
also conserved with $D=0$ for $\nu$, $D=1$ for $\eta^0,\zeta^0$, and $D=-1$ 
for $N$.  So far, the charged scalar gauge singlet $\zeta^- \sim 3'$ of 
Table~1 has not been used.  It has $L=D=1$ and allows $N$ to be produced 
as freeze-in dark matter~\cite{hjmw10} through Higgs decay~\cite{m19} as 
shown in Fig.~2.  It is the analog of Fig.~1, but now $L$ circulates in 
the loop, whereas $D$ flows from $N$ to $\zeta^-$ to $N$.  Here $A_5$ is 
unbroken by all the couplings and the one-loop diagram is finite because 
 \begin{figure}[htb]
 \vspace*{-5cm}
 \hspace*{-3cm}
 \includegraphics[scale=1.0]{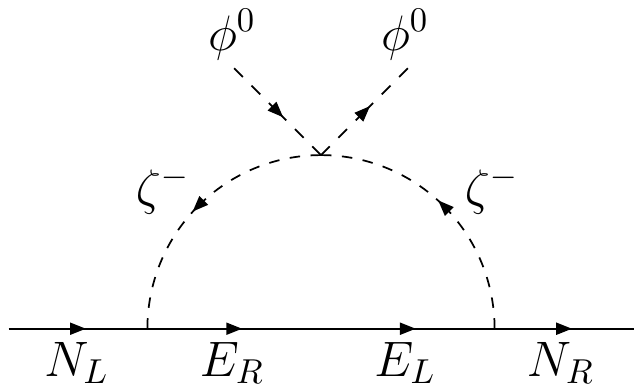}
 \vspace*{-21.5cm}
 \caption{Higgs decay to $\bar{N}N$.}
 \end{figure}
it involves three propagators, two scalar and one fermion.  It yields the 
effective Yukawa coupling $f_h$ of the Higgs boson $h$ to $\bar{N}N$.
\begin{equation}
f_h = {\lambda_\zeta f^L_\zeta f^R_\zeta v m_E \over 16 \pi^2} \left[ 
{1 \over m_\zeta^2 - m_E^2} -{m_E^2 \ln(m_\zeta^2/m_E^2) \over 
(m_\zeta^2-m_N^2)^2} \right].
\end{equation}
The decay rate of $h$ to $\bar{N}N$ is
\begin{equation}
\Gamma_h = {f_h^2 m_h \over 32 \pi} \sqrt{1-4r^2}(1-2r^2),
\end{equation}
where $r=m_N/m_h$.

As shown in Ref.~\cite{mr21}, if the reheat temperature $T_R$ of the Universe 
after inflation is below the decoupling temperature of $N$ but above $m_h$, 
say $T_R \sim 1-10$ TeV, then $N$ is a feebly interacting massive particle 
(FIMP), which only production mechanism is freeze-in, through Higgs decay, 
before the latter decouples from the thermal bath.  Typical values for 
this to happen here are $m_N \sim$ GeV, $f_h \sim 10^{-11}$, 
$m_\zeta \sim 10^4$ GeV, and $m_E \sim 10^5$ GeV.

\noindent \underline{\it Concluding Remarks}~:~
The non-Abelian discrete symmetry $A_5$ of the 4-dimensional pentatope is 
used to construct a radiative model of Dirac neutrinos through dark matter, 
with dark number $D$ derived from lepton number $L$, i.e. $D=L-2j$. 
Cobimaximal neutrino mixing is obtained from the soft breaking of $A_5$ 
to $A_4$ which is the symmetry of the 3-dimensional tetrahedron.
The complete neutrino mass matrix is linked to the dark neutral fermions, 
with a realistic numerical example showing the normal ordering of neutrino 
masses. The dark fermions are produced by the freeze-in mechanism through 
the decay of the SM Higgs boson.

\noindent \underline{\it Acknowledgement}~:~
This work was supported in part by the U.~S.~Department of Energy Grant 
No. DE-SC0008541.  

\bibliographystyle{unsrt}

\end{document}